\documentclass[12pt,a4paper,twoside]{article}
\usepackage{epsfig}
\usepackage{amsfonts}
\usepackage{amssymb}

\newcommand{\be}{\begin{equation}}
\newcommand{\ee}{\end{equation}}
\newcommand{\bd}{\begin{displaymath}}
\newcommand{\ed}{\end{displaymath}}
\newcommand{\bea}{\begin{eqnarray}}
\newcommand{\eea}{\end{eqnarray}}
\newcommand{\non}{\nonumber}

\begin{document}
\begin{flushright}
INLO-PUB-1/99\\
\end{flushright}

\vspace{.5cm}

\begin{center}
{\Large Non--trivial flat connections on the 3--torus I\\}
{$G_2$ and the orthogonal groups}

\vspace{.5cm}

{\bf Arjan Keurentjes}\footnote{email address: \sl arjan@lorentz.leidenuniv.nl} \\
{\it Instituut-Lorentz for theoretical physics, Universiteit Leiden \\ P.O. Box 9506, NL-2300 RA Leiden, 
The Netherlands}\end{center}

\vspace{.5cm}

\begin{abstract}
We propose a construction of non-trivial vacua for Yang-Mills theories on the 3--torus. Although we consider theories with periodic boundary conditions, twisted boundary conditions play an essential auxiliary role in our construction. In this article we will limit ourselves to the simplest case, based on twist in $SU(2)$ subgroups. These reproduce the recently constructed new vacua for $SO(N)$ and $G_2$ theories on the 3--torus. We show how to embed the results in the other exceptional groups $F_4$ and $E_{6,7,8}$ and how to compute the relevant unbroken subgroups. In a subsequent article we will generalise to $SU(N > 2)$ subgroups. The number of vacua found this way exactly matches the number predicted by the calculation of the Witten index in the infinite volume.    
\end{abstract}

\section{Introduction}
When constructing gauge field theories on a torus, a logical starting point is to begin with the construction of the classical vacua. For a gauge theory, this means finding all solutions with vanishing field strength, $F_{\mu \nu}= 0$, while respecting the appropriate boundary conditions. For Yang-Mills theories the relevant equation is non-linear, and hence non-trivial. With periodic boundary conditions, one trivial class of solutions is easily found, one can set $A_{\mu} = c^{a}_{\mu} H_a$ with $c^a_{\mu}$ a constant, and $H_a$ group generators in a maximal abelian algebra of the group (Cartan subalgebra (CSA)). For $SU(N)$ and $Sp(N)$, all vacuum solutions are gauge equivalent to a member of this class, but in general this turns out not to be the case.

This problem is of relevance for the calculation of the Witten index, ${\rm Tr} (-1)^F$ in 4-dimensional supersymmetric gauge theories (where $F$ stands for fermion number, and the trace is over the Hilbert space). In a classic paper \cite{IWit}, Witten proposed to formulate the theory on $T^3 \times R$ with periodic boundary conditions (For $SU(N)$, ${\rm Tr} (-1)^F$ was also computed on $T^3 \times R$ with twisted boundary conditions, but this approach cannot be applied to a general group). The expectation is, that for a sufficiently small size of the torus, a semiclassical calculation will be reliable. The energy spectrum is discrete, the vacua are distinct, and their contribution to ${\rm Tr} (-1)^F$ can in principle be counted. Based on the above mentioned trivial class of solutions, Witten predicted ${\rm Tr} (-1)^F= r+1$ for theories with a simple gauge group, where $r$ is the rank of the gauge group. It is argued that ${\rm Tr} (-1)^F$ is invariant under a large class of perturbations, and one can ask the question whether this result can be reproduced in the infinite volume limit, the theory on $R^4$. In the infinite volume, the gauge theory exhibits chiral symmetry breaking via the axial anomaly, and, assuming gluino condensation, one finds $h$ distinct vacua \cite{IWit, IDyn} ($h$ is the dual Coxeter number, which is equal to the Dynkin index of the adjoint representation, in an appropriate normalisation). The argument therefore suggests that $h = r+1$ for simple gauge groups. This equality involves two purely group-theoretical quantities, and is easily checked. One finds that the equality is satisfied for the unitary and symplectic theories, but \emph{not} for the theories based on orthogonal or exceptional gauge groups.

A resolution to this paradox was offered in \cite{Witnew}. Using an M-theory construction, Witten was able to show that in $SO(N)$ with $N \geq 7$, flat connections exist which are not of the trivial type. These non-trivial flat connections give an extra contribution to ${\rm Tr} (-1)^F$ of $r'+ 1$ where $r'$ is the rank of the subgroup that is unbroken by the holonomies (Wilson loops along the non-trivial cycles of the torus). On the basis of his analysis Witten suggested that
\be \label{Cox}
h = \sum_i (r_i + 1)
\ee
where the sum runs over disconnected components of moduli space, i.e. different classes of solutions, and $r_i$ is the rank of the unbroken gauge group in each of these components. For $SU(N)$ and $Sp(N)$, the moduli space has only one component, and (\ref{Cox}) reduces to the old formula. For the orthogonal groups, the extra vacua of Witten make up a second component in moduli space, and the result satisfies the equality (\ref{Cox}). Witten conjectured that extra solutions might also solve the problem for the exceptional groups. Indeed, for $G_2$, it was shown in \cite{Keur1} that an extra vacuum exists, and (\ref{Cox}) is satisfied. We will describe a method that allows us to address the question for the other exceptional groups.

The methods involved so far to find extra vacua rely heavily on the properties of orthogonal groups \cite{Witnew, Keur1}. The $G_2$ case is solvable because of the relation of $G_2$ to $SO(7)$. In this paper, a more systematic way of constructing non-trivial vacua will be discussed. The existence of extra vacua in $G_2$ and $SO(N)$ will be linked to the fact that these groups have special subgroups. This allows us to reconstruct the holonomies explicitly. Our construction can easily be embedded in the exceptional groups $F_4$ and $E_{6,7,8}$. In this paper we will restrict ourselves to the simplest realisation of our construction, based on twist in $SU(2)$-subgroups. This will be sufficient to reproduce the previous results from \cite{Witnew, Keur1} for $G_2$ and the orthogonal groups. In a subsequent article \cite{Keur3} we will show that constructions with twist in $SU(N)$-subgroups (with $N> 2$) can be realised in the other exceptional groups (and only in these). Although we do not know how to prove whether our method gives a complete classification of the flat connections for exceptional gauge groups on $T^3$, a strong indication that this is the case, is the fact that (\ref{Cox}) is satisfied when the new vacua are included.   

The outline of this paper will be as follows. In section \ref{math} we will introduce our methods, and some machinery. In the next section we will apply our construction to find the new vacua, and calculate the unbroken gauge group. In an appendix we briefly review Lie algebras of simple Lie groups to outline our conventions. In a subsequent article \cite{Keur3} we will construct more vacua for the exceptional groups, and demonstrate how the Witten index problem is solved. These results are summarised in two tables.   

\section{Non-trivial flat connections} \label{math}
\subsection{Holonomies and vacua}

For any flat periodic connection on a torus, one can define a set of 
holonomies, Wilson loops along nontrivial cycles of the torus,
\be \label{hol}
\Omega_k \ =\ P \exp \left\{ i \int_0^{L_k} A_k(x) dx^k \right\},
\ee
where $k=1,2,3$ labels the holonomy corresponding to the cycle wrapping around the $x,y,z$ direction respectively, $L_k$ is the length of the cycle.
 
The fact that the connection is flat will imply that the holonomies commute. If one is only working with fields that take values in a representation of the gauge group that does not faithfully represent the center (implying that the representation is not simply connected), it is possible to impose commutativity up to an element of the center \cite{Hooft}. Working in a representation of the group that is simply connected, one can easily distinguish between commutativity, and commutativity up to a center element (twisted boundary conditions). We mention these points because twisted boundary conditions will form an essential ingredient in what follows, although in a somewhat unexpected way. As shown in \cite{Keur1}, the fact that holonomies commute in a simply connected representation, is sufficient for the existence of a corresponding flat connection. Thus, we can use the holonomies (\ref{hol}) to characterise the flat connections. 

Under periodic gauge transformations, the holonomies transform covariantly ($\Omega_k' = g \Omega_k g^{-1}$), and hence their traces are gauge invariant. For the trivial class of solutions $A_{\mu} = c^a_{\mu} H_a$, the holonomies $\Omega_k = \exp (i c^a_k H_a L_k)$ are on the so-called maximal torus (obtained by exponentiating the CSA). To find a different solution, one has to find holonomies that commute, but do not lie on a maximal torus. The corresponding flat connections are no longer of a simple form in these cases.

For $T^2$, using the complex structure on the 2-torus, it is possible to prove (as sketched in a footnote in \cite{Witnew}) that the moduli space of flat connections is connected, and hence the trivial class is the only class of solutions . For $T^3$ this is not the case, but using the result for $T^2$ one can arrange that any two out of the three holonomies are exponentials of elements of the CSA. The third holonomy is therefore the crucial one: either it can be written as the exponent of an element of the CSA, and we have a trivial solution, or it cannot, and one has a non-trivial solution.

The local parameters of the moduli space can be found by perturbing the holo\-nomies $\Omega_k$ around a solution, demanding they still commute (so that the perturbations also lead to admissible vacua). An infinitesimal perturbation of the holonomy looks as follows (with $T^a$ the group generators)
\be \label{pert}
\Omega'_k = \Omega_k(1 + i\alpha_k^a T^a)
\ee
When we require that all $\Omega'_k$ still commute, this implies the conditions \cite{Keur1}
\be \label{cond}
\alpha_k^a \Omega_k [T^a, \Omega_l] = \alpha_l^a \Omega_l [T^a, \Omega_k]  
\ee
If there are generators $T^a$ that commute with each of the $\Omega_k$, this equation does not lead to any restrictions on the corresponding $\alpha_k^a$, and these generators correspond to admissible perturbations. They will generate a group which is unbroken by the holonomies, and this is the group that is relevant in the calculation of ${\rm Tr} (-1)^F$. We now show that the generators that do not commute with all holonomies correspond to global gauge transformations, for $\Omega_k$ based on the $SU(N)$ subgroup construction to be described (for $N > 2$ see ref. \cite{Keur3}). From this construction it follows that one can choose a basis $T^a$ for the Lie algebra such that
\be \label{comm}
\textrm{Ad}(\Omega_k)T^a = \Omega_k T^a \Omega_k^{-1}= \exp(i \frac{2 \pi n^a_k}{N}) T^a \quad n^a_k, N \in \mathbf{Z} \quad \forall k,a .
\ee
In that case the condition (\ref{cond}) reads
\be \label{cond2}
\alpha_k^a (\exp(-i \frac{2 \pi n^a_l}{N})-1) \Omega_k \Omega_l T^a = \alpha_l^a (\exp(-i \frac{2 \pi n^a_k}{N})-1) \Omega_k \Omega_l T^a ,  
\ee
Restricting ourselves to generators that do not commute with all holonomies, we define for $k$ such that $[\Omega_k , T^a] \neq 0$
\be
\beta^a = \frac{\alpha^a_k}{1- \exp(-i \frac{2 \pi n^a_k}{N})}.
\ee
Equation (\ref{cond2}) implies that $\beta^a$ is independent of $k$, whereas (\ref{cond}) implies $\alpha^a_k = 0$ if $[ \Omega_k, T^a ] = 0$, so we can write 
\be \label{cond2a}
i \alpha^a_k \Omega_k T^a = i\beta^a [ \Omega_k, T^a ].
\ee
We are thus left with a global gauge transformation, generalising a result derived in \cite{Keur1} for $SO(7)$.

\subsection{Method of construction}

Our construction is based on a variant of the so-called multi-twisted boundary conditions, as considered first in \cite{Cohen} involving subgroups of the gauge group.

We restrict ourselves to Yang-Mills theories with a compact, simple and simply connected gauge-group $G$. We will be interested in subgroups $PG_N'$ whose universal covering $\widetilde{PG_N}'$ is the product of $N$ factors $\tilde{G}'$. The representation of $PG_N$ will in general not be irreducible, nor are its irreducible components in the same congruence class. 

The \emph{global} structure of our subgroup $PG_N'$ will \emph{not} be that of a direct product group. For the realisation of twisted boundary conditions, it is necessary that a non-trivial discrete central subgroup has been divided out; for multi-twisted boundary conditions this discrete central subgroup is diagonal. As a relevant example of such a subgroup, consider $SO(4)$ which is locally $SU(2) \times SU(2)$, but has global structure $(SU(2) \times SU(2))/ \mathbf{Z}_2$, where $\mathbf{Z}_2$ is the diagonal subgroup in the $\mathbf{Z}_2 \times \mathbf{Z}_2$ centre of $SU(2) \times SU(2)$.

For the discussion here we will restrict ourselves to a subgroup $PG_2'$ (with universal covering $\tilde{G}'^2$), since the generalisation to $PG_N'$ will be obvious. We assume that $\tilde{G}'$ allows a non-trivial center $Z$, and that the global subgroup $PG_2'$ is a representation of $(\tilde{G}' \times \tilde{G}')/Z_{diag}$ where $Z_{diag} \cong Z$ is the diagonal subgroup of the centre $Z \times Z$ of $\tilde{G}'^2$. Now select two elements $P_1$, $Q_1$ generated by the Lie algebra of the first $\tilde{G}'$ factor ($\tilde{G}'_1$), and two elements $P_2$, $Q_2$ generated by the Lie algebra of the second $\tilde{G}'$ factor ($\tilde{G}'_2$), such that they commute up to a non-trivial element of the center $Z$ of $\tilde{G}'_i$:
\bea
P_1 Q_1 & = & z_1 Q_1 P_1 \label{com1}\\
P_2 Q_2 & = & z_2 Q_2 P_2 \label{com2}
\eea
For irreducible representations of $\tilde{G}'_i$, $z_i$ is a root of unity, for a reducible representation $z_i$ is a diagonal matrix, with on the diagonal the center elements appropriate for the different irreducible components. Since we only deal with representations of $(\tilde{G}' \times \tilde{G}')/Z_{diag}$, $z_1$ and $z_2$ are actually elements of the central subgroup $(Z \times Z)/Z_{diag} \cong Z$ . By picking $P_i$ and $Q_i$ in a specific way, we can thus arrange that (\ref{com1}, \ref{com2}) are satisfied with the additional condition 
\be \label{centercond}
z_1 = (z_2)^{-1} \equiv z
\ee
We now have
\be
P = P_1 P_2 \qquad Q = Q_1 Q_2 \quad \Rightarrow PQ = QP
\ee    

It is possible\footnote{It was proven by Dynkin \cite{Dynkin} that if a Lie algebra ${\cal L}_G$ has a subalgebra ${\cal L}_{G'}$, then the CSA of ${\cal L}_G'$ can be chosen to be contained in the CSA of ${\cal L}_G$ (by applying a suitable automorphism of ${\cal L}_G$). Upon exponentiation, one finds that the maximal torus of the group $G'$, generated by ${\cal L}_{G'}$, is contained in the maximal torus of the group $G$, generated by ${\cal L}_G$.} \label{foot} and convenient to embed the $PG_2'$ subgroup in such a way that its maximal torus $T'$ is a subgroup of a maximal torus $T$ of $G$. If the $PG_2'$-subgroup has the same rank as $G$, then the tori $T$ and $T'$ coincide. If this is not the case, then there are multiple ways to extend $T'$ to a maximal torus $T$ of $G$.

We can choose the elements $P_1$ and $P_2$ to lie on the torus $T'$. Since the maximal torus is abelian, it immediately follows that neither $Q_1$ nor $Q_2$ is on $T'$, and neither is their product $Q$ (since $Q$ does not commute with either of the $P_i$). We will now construct a third element $P'$ by ``twisting'' one of the $G'$'s with respect to the other $G'$
\be \label{Qconj}
P' = Q_1^n P Q_1^{-n} =  Q_1^n P_1 Q_1^{-n} P_2 = P_1 Q_2^{-n} P_2 Q_2^{n} = z^{-n} P 
\ee
We can define a set of holonomies by setting $\Omega_1 = P$ , $\Omega_2 = P'$ and $\Omega_3 = Q$. We will always assume $P'$ and $P$ to be different, which is essential for finding non-trivial vacua. This limits $n$ to a finite set, since there exists some $n$ for which $z^{-n} = 1$. We also should not allow $z$ to be an element of the centre of $G$, since this will also imply that the connection defined by the holonomies $P$, $P'$ and $Q$ is trivial (we can arrange that $P$ and $Q$ are on a maximal torus of $G$, and then, since the centre of $G$ is also on this maximal torus, $P'$ must be on this maximal torus).
Since $P$ is on the maximal torus $T'$, and $z$ is an element of the centre of $PG_2'$, and the center is generated by the CSA, $P'$ is on the maximal torus $T'$. By assumption the torus $T'$ is contained in a maximal torus $T$ of $G$. To define a non-trivial flat connection $Q$ should not be on any maximal torus $T$ of $G$. If $Q$ is on a maximal torus of $G$, then the maximal abelian subgroup that commutes with $P$, $P'$ and $Q$ is a maximal torus.

Since the construction involves subgroups $\tilde{G}'$ with a non-trivial center, one may always take a subgroup of $\tilde{G}'$ that is a product of unitary groups\footnote{If some simple subgroup with non-trivial center is not $SU(N)$, its center is either $\mathbf{Z}_2$ ($Sp(n)$, $SO(2n+1)$ and $E_7$), $\mathbf{Z}_4$ ($SO(4n+2)$) , $\mathbf{Z}_2 \times \mathbf{Z}_2$ ($SO(4n)$) of $\mathbf{Z}_3$ ($E_6$). In these cases the center is contained in an $SU(2)$, $SU(4)$, $SU(2) \times SU(2)$ or $SU(3)$-subgroup}. Henceforth, we shall allways assume $\tilde{G}'$ to be a product of $SU(N)$'s. Thus we may take $\tilde{G}'= SU(N_1)^{n_1} \times SU(N_2)^{n_2} \times \cdots$ ($N_i$ different, $n_i$ positive integers), which has as center $(\mathbf{Z}_{N_1})^{n_1} \times (\mathbf{Z}_{N_2})^{n_2} \times \cdots$. 

\subsection{Diagonal subgroups}

Although not strictly necessary, extremely useful for our calculations is the concept of a diagonal subgroup. It is possible to construct a diagonal subgroup $D$ (with universal covering $\tilde{D} = \tilde{G}'$) in $PG_N'$ as follows: construct a Lie algebra ${\cal L}$ for $\tilde{G}'$, consisting of elements $T^a$. Then the Lie algebra for $(PG_N'$ has the structure ${\cal L}_1 \oplus {\cal L}_2 \oplus \cdots \oplus {\cal L}_N$ with each of the ${\cal L}_i \cong {\cal L}$. Hence we can write $T^a_i$, for the generator from ${\cal L}_i$ that corresponds to $T^a$ under an isomorphism mapping ${\cal L}$ to ${\cal L}_i$. The diagonal subgroup of $PG_N'$ is then constructed by taking as generators $T^a_1 + T^a_2 + \cdots + T^a_N$. This construction is not unique, there are many isomorphisms from ${\cal L}_i$ to ${\cal L}$, and these will give different (but isomorphic) diagonal subgroups.

We now take $P_1, P_2, Q_1, Q_2$  as before. The $P_i$ are elements of the maximal torus of $\tilde{G}'_i$, so we can write $P_i = \exp (i h_i)$ with $h_i$ an element of the CSA of ${\cal L}_i$, and similarly $Q_i = \exp(i e_i)$ with $e_i$ \emph{not} in the CSA. The elements $P = P_1 P_2 = exp (i (h_1 + h_2) )$ and $Q = Q_1 Q_2 = exp (i (e_1 + e_2))$ are then elements of a diagonal group $D$, as constructed in the above. Conjugating with $Q_1^n$ will produce $Q_1^n P Q_1^{-n} = P'$ and $Q_1^n Q Q_1^{-n} = Q$, which are elements of a diagonal subgroup $D'$, isomorphic to $D$ (the isomorphism being given by $D \leftrightarrow Q_1^n D Q_1^{-n}$).  

\section{Non-trivial vacua based on twist in $SU(2)$}

In this section we specialise to $\tilde{G}'= SU(2)$. We will start by developing the relevant tools for this subgroup. After that we will give an overview of groups in which our construction can be realised. These include $SO(N)$ where the result from \cite{Witnew} is reproduced, and $G_2$, where we rederive the result from \cite{Keur1}.

\subsection{Twist in $SU(2)$}

We will use the following convention for the $su(2)$ algebra:
\be \label{su2}
[ L_3, L_+ ] = L_+ \qquad 
[ L_3, L_- ] = -L_- \qquad 
[ L_+, L_- ] = 2 L_3
\ee
We have $L_3^{\dagger} = L_3$ and $L_+^{\dagger} = L_-$. With these conventions the eigenvalues of $L_3$, $(L_+ + L_-)/2$ and $i(L_+ - L_-)/2$ are half-integers for representations that exponentiate to $SU(2)$, and integers for representations that exponentiate to $SO(3)$.

In $SU(2)$ we will be looking for elements $p$ and $q$ such that
\be
pq = -qp
\ee
The standard convention is to take:
\be 
p = \left( \begin{array}{cc} i & 0 \\ 0 & -i \end{array} \right) = \exp \frac{i \pi}{2} \left( \begin{array}{cc} 1 & 0 \\ 0 & -1 \end{array} \right) \qquad q = \left( \begin{array}{cc} 0 & i \\ i & 0 \end{array} \right)= \exp \frac{i \pi}{2} \left( \begin{array}{cc} 0 & 1 \\ 1 & 0 \end{array} \right) 
\ee
In terms of generators this is $p = \exp(i \pi L_3)$ and $q = \exp(i \pi (L_+ + L_-)/2)$ (When lifted to $SO(3)$, the elements are $p= \textrm{diag}(-1,-1,1)$ and $q= \textrm{diag}(1,-1,-1)$, which commute). The commutation relations of $p,q$ with the algebra elements are easily determined
\be
\begin{array}{rclcrcl}
p L_3 & = & L_3 p & & q L_3 & = & - L_3 q  \\
p L_+ & = & -L_+ p & & q L_+ & = & L_- q \\
p L_- & = & -L_- p & & q L_- & = & L_+ q
\end{array}
\ee
Note that $q$ induces the Weyl reflection on the root lattice. This has a nice analogon for twist in $SU(N), N > 2$ \cite{Keur3}. 

The elements $P_i$ will take the role of $p$ in the above, the elements $Q_i$ take the role of $q$. Now notice that the condition (\ref{centercond}) implies that the diagonal group $D$ that contains $P = P_1 P_2$ and $Q = Q_1 Q_2$ is actually an $SO(3)$ (it has a trivial center). This can be seen as follows: the diagonal subgroup-construction provides a homomorphism from any of the $G'\cong SU(2)$ factors to the diagonal subgroup $D$. Under this homomorphism the non-trivial center element of $SU(2)$ is mapped to the identity in $D$ (because $P_i \rightarrow P$, $Q_i \rightarrow Q$, we have $P_i Q_i P_i^{-1}Q_i^{-1} = z \rightarrow PQP^{-1}Q^{-1} = 1$). 

\subsection{Calculation of the unbroken subgroup} \label{calc}

One of the issues we did not address so far is the fact that we require the holonomies $\Omega_i$ ($P$, $P'$ and $Q$) to commute in a simply connected representation (otherwise the theorem of \cite{Keur1} might not work). In fact we will show that they commute in \emph{any} representation, which seems more general, but is equivalent to the previous statement by the theory of compact Lie groups. For the $SU(2)$-based construction described in this paper, a sufficient condition for the commuting of all holonomies is that $D$ is an $SO(3)$-group (since this will imply that $P$ and $Q$ commute, and from this it follows that $P'$ and $Q$ commute. $P$ and $P'$ commute by construction). To determine whether $D$ is an $SO(3)$ subgroup, we construct its algebra. 

As remarked in a footnote on \pageref{foot}, it is always possible to choose an embedding of a subgroup such that its CSA is contained in the CSA of the group that contains the subgroup, so let $L_3 = h_{\zeta}$ for some $h_{\zeta}$ in the CSA of $G$. The eigenvalues of $h_{\zeta}$ are determined by taking inner products with the weights of $G$. We want $L_3$ to be an $SO(3)$-generator, and therefore its eigenvalues should be integers. Therefore $\langle \lambda, \zeta \rangle$ should be integer for any weight $\lambda$. Since any weight is expressible as a linear combination of fundamental weights and simple roots with integer coefficients, the condition that $L_3$ should have integer eigenvalues can be translated to
\bea
\langle \alpha_i , \zeta \rangle & \in & \mathbf{Z} \\
\langle \Lambda_i , \zeta \rangle & \in & \mathbf{Z}
\eea
where $\alpha_i$ are the simple roots of $G$, and $\Lambda_i$ are its fundamental weights. However, if the first of these two conditions is satisfied, then the second condition is satisfied for some weights, namely those fundamental weights with $\Lambda_i = \sum_k q_k \alpha_k$ with $q_k$ integer (we call these integer weights). Hence the second condition only has to be checked for what we will call non-integer weights. A list of these is contained in our appendix.

We will want to compute products like $pTp^{-1}$, $qTq^{-1}$, where $T$ are generators of the group $G$ we are working in, and $p$ and $q$ are as in the previous paragraph. We will be looking at $SO(3)$-subgroups, and the generators of $G$ split into $SO(3)$ representations. $pTp^{-1}$ is easily calculated. With $L_3= h_{\zeta}$ we have
\be \label{Pcomm}
p h_{\alpha} p^{-1} = h_{\alpha} \qquad pe_{\alpha}p^{-1} = \exp(i \pi\langle \alpha, \zeta \rangle ) e_{\alpha}
\ee
Note that this implies that the generators either commmute or anticommute with $p$.

There is also a nice and easy way to compute $qTq^{-1}$. First decompose the representation of $G$ into irreducible representations of $SO(3)$. In each irrep, construct the normalised eigenvectors $\psi_{\lambda}$ of $L_3$: $L_3 \psi_{\lambda} = \lambda \psi_{\lambda}$. It then follows that
\be 
L_3 q \psi_{\lambda} = - q L_3 \psi_{\lambda} = - \lambda q \psi_{\lambda}
\ee
Hence we conclude $q \psi_{\lambda} = \phi_{\lambda} \psi_{-\lambda}$ (where $ \phi_{\lambda}$ is a phase factor). In fact, since we are only dealing with $SO(3)$ representations of $q$, we know that the eigenvalues of $q$ should be $\pm 1$ and hence $q^2 =1$, meaning that $\phi_{\lambda} \phi_{-\lambda} = 1$. Moreover from $q L_+ = L_- q$ we easily find that\footnote{We use the Condon-Shortley phase convention: $L_+ \psi^j_{\lambda}= \sqrt{(j - \lambda)(j + \lambda +1)} \psi^j_{\lambda+1}$ and $L_- \psi^j_{\lambda}= \sqrt{(j + \lambda)(j - \lambda +1)} \psi^j_{\lambda-1}$, where $j$ is the usual angular momentum number, related to the dimension $d$ of the representation by $j = (d-1)/2$}
\be
\phi_{\lambda} \psi_{-(\lambda+1)} = \phi_{\lambda+1} \psi_{-(\lambda+1)}
\ee   
So, $\phi_{\lambda}= \phi_{\lambda+1} = \phi$, independent of $\lambda$, and $\phi^2 = 1$. It is now trivial to construct eigenvectors and eigenvalues for $q$:
\bea
\psi_{\lambda} + \psi_{-\lambda} & \rightarrow & \textrm{eigenvalue }\phi \non \\
\psi_{\lambda} - \psi_{-\lambda} & \rightarrow & \textrm{eigenvalue }-\phi
\eea 
Now remember once more that only $SO(3)$ representations occur, meaning that $\lambda \in \{ 0 , \pm 1, \pm 2, \ldots \}$. If the dimension of our representation is $2n+1$, then the eigenvalue  $\phi$ occurs $n+1$ times, and  $-\phi$ $n$ times. The determinant of the matrix is then $(-)^n \phi^{2n+1} = (-)^n \phi$. The determinant should however be $1$ since the matrix has been obtained by exponentiating a traceless generator. Hence we find $\phi = (-)^n$, and the action of $q$ on any representation of $SO(3)$ is fully determined, and since by assumption $G$ splits into $SO(3)$ irreps, the action of $q$ in $G$ is completely determined. Most of the above is also valid for $SU(2)$-irreps, but there are two differences: one finds $\phi^2 = -1$, and it is not possible to determine whether $\phi = i$ or $-i$. This ambiguity comes from the center of $SU(2)$, which we are unable to detect since we are trying to determine $q$ from commutation relations alone.
Notice that these considerations imply that for an appropriately chosen basis of generators $T$ of $G$ that (compare to (\ref{comm}))
\be
p T^a p^{-1} = \pm T^a \quad q T^a q^{-1} = \pm T^a
\ee

\subsection{Realisations of the $SU(2)$-based construction}

We will now discuss the cases in which our construction actually gives a non-trivial flat connection. Our conventions concerning roots and weights can be found in our appendices. For the decomposition of groups into subgroups, use was made of \cite{McKay}.

Although in each case our construction can be carried out in a subgroup $PG_2' = SO(4)$ , we will often take $PG'_N$ with $\widetilde{PG_N'} = SU(2)^N$ where $N > 2$. This allows us to choose $PG_N'$ to be a regular subgroup \cite{Dynkin}, that is, the root-lattice of $PG_N'$ is a sublattice of the root-lattice of $G$. This gives an enormous simplification of the calculations. Our methods are not limited to regular subgroups, and we have actually carried out our construction for several irregular embeddings, but we always found the same results as for the regular embeddings. Therefore we will describe only constructions with regular embeddings.

\subsubsection{$G_2$}

Our first example will be the non-trivial flat connection in $G_2$, described in \cite{Keur1}. We will treat this example in full detail to clarify our methods.
$G_2$, being the group of lowest rank that posseses non-trivial flat connections on $T^3$, is the simplest from the point of view of our construction (unlike the constructions in \cite{Witnew, Keur1} that are simpler for orthogonal groups).

\begin{figure}[!ht]
\begin{center}
\includegraphics[width=7cm]{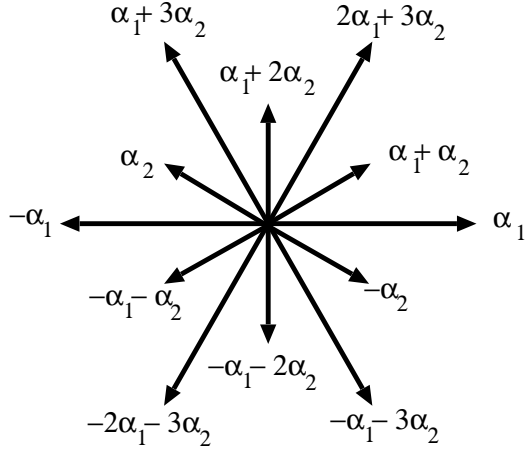}
\end{center}
\caption{the root diagram for $G_2$}
\end{figure}

$G_2$ posseses an $SO(4) \cong (SU(2) \times SU(2))/\mathbf{Z}_2$ subgroup. The first $SU(2)$-factor can be taken to be generated by $h_{\alpha_1}, e_{\alpha_1}$ and $e_{-\alpha_1}$, the second one is then generated by $h_{\alpha_1 + 2 \alpha_2}, e_{\alpha_1 + 2 \alpha_2}$ and $e_{-(\alpha_1 + 2 \alpha_2)}$. We label the two $SU(2)$ factors $SU(2)_{1,2}$, and normalise their generators such that they satisfy the algebra (\ref{su2}):
\be
\begin{array}{rcl}
SU(2)_1 & \times & SU(2)_2 \\

l_3^1 = 4 h_{\alpha_1} & & l_3^2 = 12 h_{\alpha_1 + 2 \alpha_2} \\
l_+^1 = 2 e_{\alpha_1} & & l_+^2 = 2 \sqrt{3} e_{\alpha_1 + 2 \alpha_2}
\end{array}
\ee
The diagonal subgroup is now easily constructed, being generated by 
\bea
L_3^D = & l_3^1 + l_3^2 & = h_{16 \alpha_1 + 24 \alpha_2} \\
L_+^D = & l_+^1 + l_+^2 & = 2 e_{\alpha_1} + 2 \sqrt{3} e_{\alpha_1 + 2 \alpha_2} \non
\eea
The diagonal $SU(2)$ turns out to be an $SO(3)$. To check this we construct the weights for the $su(2)$-representation. These are $\langle 16 \alpha_1 + 24 \alpha_2, \lambda \rangle$ where $\lambda$ are the weights of $G_2$. However, since any weight of $G_2$ is of the form $p \alpha_1 + q \alpha_2$, with $p,q$ integer, it is sufficient to compute $\langle 16 \alpha_1 + 24 \alpha_2, \alpha_1 \rangle = 1$ and $\langle 16 \alpha_1 + 24 \alpha_2, \alpha_2 \rangle = 0$. We find that the weights of the $su(2)$ are always integer, no matter what representation of $G_2$ we use, and hence the $su(2)$ is actually an $so(3)$. We find the following decompositions for the fundamental and adjoint irreps: 
\bea
G_2 \quad \rightarrow & SU(2) \times SU(2) & \rightarrow \quad SU(2) / \mathbf{Z}_2 \cong SO(3) \non \\
\mathbf{7} \quad \rightarrow & \mathbf{(1,3) \oplus (2,2)} & \rightarrow \quad \mathbf{3 \oplus 3 \oplus 1}\\
\mathbf{14} \quad \rightarrow & \mathbf{(3,1) \oplus (2,4) \oplus (1,3)} & \rightarrow \quad \mathbf{ 3 \oplus 5 \oplus 3 \oplus 3 } \non 
\eea
It is easy to construct the elements $P$ and $Q$: Take
\be
P = \exp (i \pi L_3^D) \quad Q = \exp(i \frac{\pi}{2} (L_+^D + L_-^D))
\ee
Now we wish to obtain $P'$. Conjugation with $Q_1 = \exp (i \frac{\pi}{2}( l_+^1 + l_-^1))$ generates the Weyl reflection in  the first of the two $SU(2)$-factors:
\be
l_3^1 \leftrightarrow -l_3^1 \qquad l_+^1 \leftrightarrow l_-^1
\ee
This will take the diagonal subgroup $D$ to a diagonal subgroup $D'$ generated by:
\bea
L_3^{D'} = & -l_3^1 + l_3^2 & = h_{8 \alpha_1 + 24 \alpha_2} \\
L_+^{D'} = & l_-^1 + l_+^2 & = 2 e_{-\alpha_1} + 2 \sqrt{3} e_{\alpha_1 + 2 \alpha_2} \non
\eea
Since $L_+^{D'} + L_-^{D'} = L_+^D + L_-^D$, $Q$ is also an element of $D'$. Constructing $P'$ proceeds as in the above:
\be
P'= \exp (i \pi L_3^{D'}) 
\ee
$P$, $P'$ and $Q$ commute by construction. It is also clear that the flat connection implied by $\Omega_1= P$, $\Omega_2=P'$ and $\Omega_3= Q$ is non-trivial, since the maximal torus of $G_2$ is simply the direct product of the tori of $D$ and $D'$, and $Q$ is not on either one.

To calculate the unbroken subgroup, we calculate the commutators of generators with the $\Omega_i$'s. As explained before only generators that commute with all three $\Omega_i$'s will be relevant. Therefore the most efficient way to proceed is to first compute the commutators with $\Omega_1=P$ and $\Omega_2=P'$, since these are the easiest, and only compute the commutator with $\Omega_3=Q$ for those generators that commute with both $P$ and $P'$.

We can use the results of section \ref{calc}. If in this section we substitute $L_3^D$ for $L_3$, and $L_{\pm}^D$ for $L_{\pm}$, then it is clear we should identify $P$ with $p$, and $Q$ with $q$. If in section \ref{calc} we substitute $L_3^{D'}$ for $L_3$, and $L_{\pm}^{D'}$ for $L_{\pm}$, then we should identify $P'$ with $p$, and $Q$ with $q$. For the commutators of the algebra with $P$, we use (\ref{Pcomm}), with $\zeta = 16 \alpha_1 + 24 \alpha_2$. We find that the CSA commutes with $P$, and $e_{\beta}$ commutes with $P$ only for $\beta = \pm \alpha_2$ and $\beta = \pm (2 \alpha_1 + 3 \alpha_2)$. For the commutators of the algebra with $P'$, we use (\ref{Pcomm}) with $\zeta = 8 \alpha_1 + 24 \alpha_2$, from which it follows that $e_{\beta}$ with $\beta = \pm \alpha_2$ and $\beta = \pm (2 \alpha_1 + 3 \alpha_2)$ anticommute with $P'$. Hence the only generators that commute with both $P$ and $P'$ are the CSA-generators $h_\beta$. 

To determine the effect of conjugation with $Q$ on these, we study the branching of $G_2 \rightarrow SO(3)$, where the $SO(3)$ is either $D$ or $D'$. We will take $D$, so the ladder operators are $L_{\pm}^D$ and the CSA-generator is $L_3^D$. We work with the generators of $G_2$ themselves, and hence we are in the adjoint representation of $G_2$. Eigenvectors of $L_3^D$ are easily found, since  $\textrm{ad}(L_3^D)e_{\beta} = [ h_{16 \alpha_1 + 24 \alpha_2}, e_{\beta} ] = \langle 16 \alpha_1 + 24 \alpha_2, \beta \rangle e_{\beta}$ in our conventions, and , $\textrm{ad}(L_3^D)h_{\beta} = 0$, $h_{\beta}$ and $e_{\beta}$ are eigenvectors of $L_3^D$. We can now split the representation in irreducible components, by the standard procedure of looking for highest eigenvalues, and then applying the ladder operators $L_{\pm}^D$ to complete a representation, constructing the orthogonal complement etc.. We find that both CSA-generators have eigenvalue $0$ in a $\mathbf{3}$ irrep of $SO(3)$. We should thus identify each CSA-generator to $\psi_0$ in a 3-dimensional representation of $SO(3)$, and using the results of section \ref{calc} we find that, since $Q \psi_0 = - \psi_0$, $Q h_{\beta} Q^{-1} = \textrm{Ad}(Q) h_{\beta} = -h_{\beta}$. The CSA generators thus anticommute with $Q$. Thus there is no generator that commutes with all $\Omega_i$, and, as already established in \cite{Keur1}, the vacuum implied by these holonomies is isolated, and there is only a discrete unbroken subgroup.

Finally we will give a matrix representation of the holonomies. The $\mathbf{7}$ irrep of $G_2$ has as its weights $0, \alpha_1 + \alpha_2, \alpha_1 + 2 \alpha_2, \alpha_2,-(\alpha_1 + \alpha_2), -(\alpha_1 + 2\alpha_2), -\alpha_2$. Using this ordering of weights, we find:
\bea
\Omega_1 = P & = & \textrm{diag}(1,-1,-1,1,-1,-1,1) \non \\
\Omega_2 = P'& = & \textrm{diag}(1,1,-1,-1,1,-1,-1) \\
\Omega_3 = Q & = & \left( \begin{array}{rrrrrrr}
-1 & 0 & 0 & 0 & 0 & 0 & 0 \\
 0 & 0 & 0 & 0 &-1 & 0 & 0 \\
 0 & 0 & 0 & 0 & 0 &-1 & 0 \\
 0 & 0 & 0 & 0 & 0 & 0 &-1 \\
 0 &-1 & 0 & 0 & 0 & 0 & 0 \\
 0 & 0 &-1 & 0 & 0 & 0 & 0 \\
 0 & 0 & 0 &-1 & 0 & 0 & 0  \end{array} \right ) \non
\eea
These can be brought to the form used in \cite{Keur1}. 

\subsubsection{$SO(7)$}

In $SO(7)$, there exists a subgroup $PG_3'$ with $\widetilde{PG_3'} = SU(2)^3$. The $SU(2)$-factors can be taken to be generated by subalgebras with elements  $h_{\beta_i}, e_{\pm \beta_i}$, with $\beta_1 = (1,-1,0)/\sqrt{10}$ for the first factor, $\beta_2 = (1,1,0)/\sqrt{10}$ for the second factor, and $\beta_3 = (0,0,1)/\sqrt{10}$ for the third. The diagonal subgroup is then (appropriate normalisations for each generator included)
\bea
L_3^D = & l_3^1 + l_3^2 + l_3^3 & = \sqrt{10} h_{(1,0,1)} \\
L_+^D = & l_+^1 + l_+^2 + l_+^3 & = \sqrt{5} e_{(1,-1,0)/\sqrt{10}} + \sqrt{5}e_{(1,1,0)/\sqrt{10}} + \sqrt{10} e_{(0,0,1)/\sqrt{10}} \non  
\eea
Again this is always an $so(3)$-algebra, which can be easily verified by calculating the inner products of $\sqrt{10} (1,0,-1)$ with the simple roots and non-integer weights. Vector, spin, and adjoint representation branch as follows:
\bea
SO(7) \quad \rightarrow & (SU(2))^3 & \rightarrow \quad SO(3) \non \\
\mathbf{7} \quad \rightarrow & \mathbf{(1,1,3) \oplus (2,2,1)} & \rightarrow \quad 2 \mathbf{(3)} \oplus \mathbf{1} \non \\
\mathbf{8} \quad \rightarrow & \mathbf{(1,2,2) \oplus (2,1,2)} & \rightarrow \quad 2 \mathbf{(3)} \oplus 2 \mathbf{(1)} \\
\mathbf{21} \quad \rightarrow & \mathbf{(3,1,1) \oplus (1,3,1) } &  \non \\
& \quad \mathbf{\oplus (1,1,3) \oplus (2,2,3)} & \rightarrow \quad \mathbf{5} \oplus 5 \mathbf{(3)} \oplus \mathbf{1}\non
\eea
To construct $D'$, twist the first factor with respect to the other two
\bd
l_3^1 \leftrightarrow -l_3^1 \qquad l_+^1 \leftrightarrow l_-^1
\ed
which leads to 
\bea
L_3^{D'} = & l_3^1 + l_3^2 + l_3^3 & = \sqrt{10} h_{(0,1,1)} \\
L_+^{D'} = & l_+^1 + l_+^2 + l_+^3 & = \sqrt{5} e_{-(1,-1,0)/\sqrt{10}} + \sqrt{5}e_{(1,1,0)/\sqrt{10}} + \sqrt{10} e_{(0,0,1)/\sqrt{10}}  \non
\eea
The set of holonomies is
\be \label{hol2}
\Omega_1 = P = \exp (i \pi L_3^D) \quad \Omega_2 = P'= \exp (i \pi L_3^{D'}) \quad \Omega_3 = Q = \exp(i \frac{\pi}{2} (L_+^D + L_-^D)) 
\ee
In the vector representation, these are equivalent to the holonomies of Witten \cite{Witnew}. There is no generator of the algebra that commutes with all three $\Omega_i$.

Two remarks are in place here. First, one might think that it is arbitrary which of the $SU(2)$-factors one chooses to twist (in the sense of eq. (\ref{Qconj})). Indeed, twisting the first factor leaving the other two the same, as in the above, will give a result equivalent to twisting the second factor while leaving the other two. However, twisting the third factor with respect to the other two will not work: If one tries
\bd
l_3^3 \leftrightarrow -l_3^3 \qquad l_+^3 \leftrightarrow l_-^3,
\ed
one finds
\bd
P' = \exp( i \pi \sqrt{10} h_{(1,0,-1)}) = \exp (i \pi \sqrt{10} h_{(0,0,-2)}) \ P
\ed
But, noticing that $\sqrt{10} h_{(0,0,1)}$ is a (correctly normalised) generator of an integral $su(2)$-subalgebra of $so(7)$, we easily find $\exp (i \pi \sqrt{10} h_{(0,0,-2)}) = \pm 1$ ($+1$ if the $so(7)$-algebra is in the same congruence class as the vector or adjoint representation, $-1$ if the $so(7)$-irrep is isomorphic to the spin representation). Hence $P$ and $P'$ differ only by an element of the centre of $Spin(7)$, and as explained in section 2.2, the connection implied by setting $\Omega_1 = P$, $\Omega_2 = P'$ and $\Omega_3 = Q$ is trivial. 

\begin{figure}[!ht]
\begin{center}
\includegraphics[width=7cm]{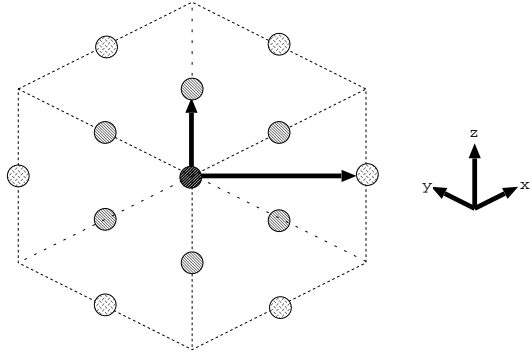}
\end{center}
\caption{projection of the roots of $so(7)$ onto those for $g_2$}
\end{figure}
As a second remark, we consider the inbedding of $G_2$ in $SO(7)$. The $so(7)$-root diagram fits into a cube. To see its $g_2$-subalgebra, project onto the plane orthogonal to a diagonal. Take as diagonal the direction $(-1,-1,1)$. The roots $\beta_2 = (1,1,0)/\sqrt{10}$ and $\beta_3 = (0,0,1)/\sqrt{10}$ will coincide under this projection (they will both project to $(1,1,2)/3\sqrt{10}$), and we find that the $SO(7)$ result can be understood from the $G_2$-result. This will be less relevant for the orthogonal groups, but important for an understanding of the exceptional groups. 

\subsubsection{$SO(8)$}

In $SO(8)$, there is a subgroup $PG_4'$, with $\widetilde{PG_4'} = SU(2)^4$. All $SU(2)$-factors can be taken to be generated by algebras with elements $h_{\beta_i}, e_{\pm \beta_i}$, with $\beta_1 = (1,-1,0,0)/\sqrt{12}$, $\beta_2 = (1,1,0,0)/\sqrt{12}$, $\beta_3 = (0,0,1,-1)/\sqrt{12}$ and $\beta_4 = (0,0,1,1)/\sqrt{12}$. The diagonal subgroup is then
\bea
L_3^D & = & \sqrt{12} h_{(1,0,1,0)} \\
L_+^D & = & \sqrt{6}(e_{(1,-1,0,0)/\sqrt{12}} + e_{(1,1,0,0)/\sqrt{12}} + e_{(0,0,1,-1)/\sqrt{12}} + e_{(0,0,1,1)/\sqrt{12}})  \non 
\eea
It is easy to verify that this is an $so(3)$-algebra.
Vector, and adjoint representation branch as follows:
\bea
SO(8) \quad \rightarrow & (SU(2))^4 & \rightarrow \quad SO(3) \non \\
\mathbf{8_v} \quad \rightarrow & \mathbf{(2,2,1,1) \oplus (1,1,2,2)} & \rightarrow \quad 2 \mathbf{(3)} \oplus 2 \mathbf{(1)} \\
\mathbf{28} \quad \rightarrow & \mathbf{(3,1,1,1) \oplus (1,3,1,1) \oplus (1,1,3,1)} &  \non \\
& \quad \mathbf{\oplus (1,1,1,3) \oplus (2,2,2,2)} & \rightarrow \quad \mathbf{(5)} \oplus 7 \mathbf{(3)} \oplus 2 \mathbf{(1)}\non
\eea
Because of triality, the spin-representations $\mathbf{8_c}$ and $\mathbf{8_s}$ have a branching that differs only from the one for $\mathbf{8_v}$ in $SU(2)^4$ by permutations of the $\mathbf{2}$'s and $\mathbf{1}$'s, their $SO(3)$-content is the same. Twist the first factor to construct $D'$:
\bea
L_3^{D'} & = & \sqrt{12} h_{(0,1,1,0)} \\
L_+^{D'} & = & \sqrt{6}(e_{-(1,-1,0,0)/\sqrt{12}} + e_{(1,1,0,0)/\sqrt{12}} + e_{(0,0,1,-1)/\sqrt{12}} + e_{(0,0,1,1)/\sqrt{12}}) \non 
\eea
The holonomies are then constructed in the usual way (\ref{hol2}). The unbroken subgroup is again discrete. The projection of $SO(8)$ onto $SO(7)$ is trivial, and we find that we can understand the $SO(8)$-result from the $G_2$-result.

\subsubsection{$SO(N)$, $N > 8$}

It is clear that the $SO(7)$-example and the $SO(8)$-example can both be embedded in $SO(9)$. Notice however, that both in the embedding of the $SO(7)$-example, and the embedding of the $SO(8)$-example we find
\bd 
L_3^D = \sqrt{12}h_{(1,0,1,0)} \quad L_3^{D'} = \sqrt{12}h_{(0,1,1,0)}
\ed
The vectors $\sqrt{12}(1,0,1,0)$ and $\sqrt{12}(0,1,1,0)$ are called ``defining vectors'' for $D$ and $D'$, and there is a theorem by Dynkin \cite{Dynkin}, that two representations of $SO(3)$ are equivalent, if their defining vectors are equivalent. Hence the $SO(7)$ and $SO(8)$ embeddings are equivalent, and will not lead to different results. One finds a $U(1) \cong SO(2)$ unbroken subgroup, but, as explained in \cite{Witnew, Keur1}, the actual unbroken subgroup is $O(2)$, because of discrete symmetries that are invisible in our approach.

In a similar way, the non-trivial flat connections for $SO(N)$ with $N > 9$ can be constructed. For $N$ sufficiently large, there are multiple ways of embedding an $SU(2)^M$ with a diagonal $SO(3)$. Although we know no simple way of proving this in our approach (their defining vectors need not be equivalent, for example), the analysis of \cite{Witnew} shows that these can never lead to new results, other than an $SO(7)$ or $SO(8)$ embedding will do. Note the chain of subgroups
\be
SO(N) \rightarrow SO(N-1) \rightarrow \cdots \rightarrow SO(7) \rightarrow G_2
\ee
that shows that non-trivial flat connections in $SO(N)$ can be derived from the $SO(N)$-subgroup $G_2$. The connected component of the maximal unbroken subgroup is $SO(N-7)$, as can be understood from the fact that $SO(N)$ branches into $G_2 \times SO(N-7)$.

\subsubsection{$F_4$}

The easiest way to proceed in $F_4$ is by using the fact that the $so(7)$ root lattice is a sublattice of the $f_4$ root lattice. We can use $su(2)$-algebra's with elements $h_{\beta_i}, e_{\pm \beta_i}$, with $\beta_1 = (1,-1,0,0)/\sqrt{18}$, $\beta_2 = (1,1,0,0)/\sqrt{18}$, and $\beta_3 = (0,0,1,0)/\sqrt{18}$. The two diagonal subgroups and holonomies are constructed in the standard way.
\bea
L_3^D = & l_3^1 + l_3^2 + l_3^3 & = \sqrt{18} h_{(1,0,1,0)} \\
L_+^D = & l_+^1 + l_+^2 + l_+^3 & = 3 e_{(1,-1,0,0)/\sqrt{18}} + 3e_{(1,1,0,0)/\sqrt{18}} + \sqrt{18} e_{(0,0,1,0)/\sqrt{18}} \non  
\eea
\bea
L_3^{D'} = & l_3^1 + l_3^2 + l_3^3 & = \sqrt{18} h_{(0,1,1,0)} \\
L_+^{D'} = & l_+^1 + l_+^2 + l_+^3 & = 3 e_{-(1,-1,0,0)/\sqrt{18}} + 3e_{(1,1,0,0)/\sqrt{18}} + \sqrt{18} e_{(0,0,1,0)/\sqrt{18}}  \non
\eea
Again these are found to be $so(3)$-algebra's. The holonomies are as in (\ref{hol2}). Calculating the (connected component of the) unbroken subgroup, one finds that it is $SO(3)$. This can be understood from the branching of $F_4$ into $G_2 \times SO(3)$. 
\bea
F_4 & \rightarrow & G_2 \times SO(3) \non \\
\mathbf{26} & \rightarrow & \mathbf{(7,3) \oplus (1,5)} \\
\mathbf{52} & \rightarrow & \mathbf{(14,1) \oplus (7,5) \oplus (1,3)} \non 
\eea
Note that here it is important that our construction fits into $G_2$. Starting from $SO(7)$ might lead to the wrong expectation that the unbroken subgroup would be $SO(2)$, since $F_4$ branches into $SO(7) \times SO(2)$. 

\subsubsection{$E_6$}

In $E_6$ we use that the $so(8)$ root lattice is a sublattice of the $e_6$ root lattice. We use $su(2)$-algebra's with elements $h_{\beta_i}, e_{\pm \beta_i}$, with $\beta_1 = (0,1,-1,0,0)/\sqrt{24}$, $\beta_2 = (0,1,1,0,0,0)/\sqrt{24}$, and $\beta_3 = (0,0,0,1,-1,0)/\sqrt{24}$, and $\beta_4 = (0,0,0,1,1,0)/\sqrt{24}$ . The two diagonal subgroups are:
\bea
L_3^D = & l_3^1 + l_3^2 + l_3^3 & = \sqrt{24} h_{(0,1,0,1,0,0)} \\
L_+^D = & l_+^1 + l_+^2 + l_+^3 & = \sqrt{12} (e_{(0,1,-1,0,0,0)/\sqrt{24}} + e_{(0,1,1,0,0,0)/\sqrt{24}} + \non \\
& & \qquad \qquad e_{(0,0,0,1,-1,0)/\sqrt{24}} + e_{(0,0,0,1,1,0)/\sqrt{24}} \non  
\eea
\bea
L_3^{D'} = & l_3^1 + l_3^2 + l_3^3 & = \sqrt{24} h_{(0,0,1,1,0,0)} \\
L_+^D = & l_+^1 + l_+^2 + l_+^3 & = \sqrt{12} (e_{-(0,1,-1,0,0,0)/\sqrt{24}} + e_{(0,1,1,0,0,0)/\sqrt{24}} + \non \\
& & \qquad \qquad e_{(0,0,0,1,-1,0)/\sqrt{24}} + e_{(0,0,0,1,1,0)/\sqrt{24}} \non  
\eea
These are $so(3)$-algebra's. The holonomies are as in (\ref{hol2}). The (connected component of the) unbroken subgroup is found to be $SU(3)$. This can be understood from the branching of $E_6$ into $G_2 \times SU(3)$. 
\bea
E_6 & \rightarrow & G_2 \times SU(3) \non \\
\mathbf{27} & \rightarrow & \mathbf{(7,\bar{3}) \oplus (1,6)} \\
\mathbf{78} & \rightarrow & \mathbf{(14,1) \oplus (7,8) \oplus (1,8)} \non 
\eea

\subsubsection{$E_7$}

The $E_6$-example can be trivially embedded in $E_7$, using that the $e_6$ root lattice is a sublattice of the $e_7$ root lattice. We add a zero to the vectors and adapt the normalisations.
\bea
L_3^D = & l_3^1 + l_3^2 + l_3^3 & = 6 h_{(0,1,0,1,0,0,0)} \\
L_+^D = & l_+^1 + l_+^2 + l_+^3 & = \sqrt{18} (e_{(0,1,-1,0,0,0,0)/6} + e_{(0,1,1,0,0,0,0)/6} + \non \\
& & \qquad \qquad e_{(0,0,0,1,-1,0,0)/6} + e_{(0,0,0,1,1,0,0)/6} \non  
\eea
\bea
L_3^{D'} = & l_3^1 + l_3^2 + l_3^3 & = 6 h_{(0,0,1,1,0,0)} \\
L_+^D = & l_+^1 + l_+^2 + l_+^3 & = \sqrt{18} (e_{-(0,1,-1,0,0,0,0)/6} + e_{(0,1,1,0,0,0,0)/6} + \non \\
& & \qquad \qquad e_{(0,0,0,1,-1,0,0)/6} + e_{(0,0,0,1,1,0,0)/6} \non  
\eea
These are $so(3)$-algebra's. The holonomies are as in (\ref{hol2}). The (connected component of the) unbroken subgroup is a simple group of 21 generators. It takes a little more work to show that it is\footnote{Some would denote what we call $Sp(n)$ as $Sp(2n)$. For our conventions e.g.$Sp(1) \cong SU(2)$, $Sp(2) /\mathbf{Z_2} \cong SO(5)$} $Sp(3)$ (and not $SO(7)$). This can be understood from the branching of $E_7$ into\footnote{In the decompositions given, both 14-dimensional representations of $Sp(3)$ are present. By $\mathbf{14}$, we denote the representation with Dynkin labels $(001)$, while $\mathbf{14'}$ is the irrep with Dynkin labels $(010)$ (the simple roots of $Sp(3)$ are ordered such that the longest one appears on the right) \cite{McKay}.}  $G_2 \times Sp(3)$
\bea
E_7 & \rightarrow & G_2 \times Sp(3) \non \\
\mathbf{56} & \rightarrow & \mathbf{(7,6) \oplus (1,14)} \\
\mathbf{133} & \rightarrow & \mathbf{(14,1) \oplus (7,14') \oplus (1,21)} \non 
\eea

\subsubsection{$E_8$}

The $e_8$ root lattice contains as a sublattice the $e_7$ root lattice. Again we add a zero to the vectors and adapt the normalisations.
\bea
L_3^D = & l_3^1 + l_3^2 + l_3^3 & = \sqrt{60} h_{(0,1,0,1,0,0,0,0)} \\
L_+^D = & l_+^1 + l_+^2 + l_+^3 & = \sqrt{30} (e_{(0,1,-1,0,0,0,0,0)/\sqrt{60}} + e_{(0,1,1,0,0,0,0,0)/\sqrt{60}} + \non \\
& & \qquad \qquad e_{(0,0,0,1,-1,0,0,0)/\sqrt{60}} + e_{(0,0,0,1,1,0,0,0)/\sqrt{60}} \non  
\eea
\bea
L_3^{D'} = & l_3^1 + l_3^2 + l_3^3 & = \sqrt{60} h_{(0,0,1,1,0,0,0)} \\
L_+^D = & l_+^1 + l_+^2 + l_+^3 & = \sqrt{30} (e_{-(0,1,-1,0,0,0,0,0)/\sqrt{60}} + e_{(0,1,1,0,0,0,0,0)/\sqrt{60}} + \non \\
& & \qquad \qquad e_{(0,0,0,1,-1,0,0,0)/\sqrt{60}} + e_{(0,0,0,1,1,0,0,0)/\sqrt{60}} \non  
\eea
These are $so(3)$-algebra's. The holonomies are as in (\ref{hol2}). The (connected component of the) unbroken subgroup is $F_4$. This can be understood from the branching of $E_8$ into $G_2 \times F_4$. 
\bea
E_8 & \rightarrow & G_2 \times F_4 \non \\
\mathbf{248} & \rightarrow & \mathbf{(14,1) \oplus (7,26) \oplus (1,52)}
\eea

\section{Non-trivial vacua based on twist in $SU(N > 2)$}

\begin{table}[p]
\begin{center}
\begin{tabular}{|c | cccccc|}
\hline
Group & & \multicolumn{4}{c}{Vacuum-type} & \\
$G$ & 1 & 2 & 3 & 4 & 5 & 6 \\
\hline
$SU(N)$    & $SU(N)$ & & & & & \\
$Sp(N)$    & $Sp(N)$ & & & & & \\
$SO(2N+1)$ & $SO(2N+1)$ & $SO(2N-6)$ & & & & \\
$SO(2N)$   & $SO(2N)$ & $SO(2N-7)$ & & & & \\
$G_2$      & $G_2$ & discrete & & & & \\
$F_4$      & $F_4$ & $SO(3)$ & discrete & & & \\
$E_6$      & $E_6$ & $SU(3)$ & discrete & & & \\
$E_7$      & $E_7$ & $Sp(3)$ & $SU(2)$ & discrete & & \\
$E_8$      & $E_8$ & $F_4$ & $G_2$ & SU(2) & discrete & discrete  \\
\hline
\end{tabular}
\caption{The connected part of the maximal unbroken subgroups} \label{tab1}
\end{center}
\end{table}

\begin{table}[p]
\begin{center}
\begin{tabular}{|c | c |cccccc|}
\hline
Group & & &\multicolumn{4}{c}{Vacuum-type} & \\
$G$ & $h$ & 1 & 2 & 3 & 4 & 5 & 6\\
\hline
$SU(N)$    & $N$   & $N$ & & & & & \\
$Sp(N)$    & $N+1$ & $N+1$ & & & & & \\
$SO(2N+1)$ & $2N-1$ & $N+1$ & $N-2$ & & & &  \\
$SO(2N)$   & $2N-2$ & $N+1$ & $N-3$ & & & & \\
$G_2$      & 4  & 3 & 1 & & & & \\
$F_4$      & 9  & 5 & 2 & (1+1) & & & \\
$E_6$      & 12 & 7 & 3 & (1+1) & & & \\
$E_7$      & 18 & 8 & 4 & (2+2) & (1+1) & & \\
$E_8$      & 30 & 9 & 5 & (3+3) & (2+2) & (1+1+1+1) & (1+1) \\
\hline
\end{tabular}
\caption{Contributions to ${\rm Tr} (-1)^F$} \label{tab2}
\end{center}
\end{table}

Although we will publish the computational details in a separate article \cite{Keur3}, we mention that our construction, with $\tilde{G}'= SU(3), SU(4), SU(5)$ and $SU(2) \times SU(3)$ can be used for the exceptional groups other than $G_2$. A construction based on $SU(3)$ is possible in $F_4$ and $E_{6,7,8}$, the $SU(4)$ construction is possible in $E_{7,8}$, and $E_8$ allows an $SU(5)$ based construction. Finally, using the $G_2 \times F_4$ subgroup of $E_8$, the $SU(2)$ and $SU(3)$ constructions can be embedded simultaneously in this group. The $SU(3)$-construction will yield two inequivalent vacuum components, obtained by taking $n$ in eq. (\ref{Qconj}) 1 or 2 respectively. The $SU(4)$ construction yields three different vacuum components, but, one of these vacuum components is already contained in the $SU(2)$-construction described in this paper. The $SU(5)$-construction gives four different vacuum components. $SU(2) \times SU(3)$ gives five different vacuum components, of which one is contained in our $SU(2)$ construction, and two are contained in our $SU(3)$-construction, so only two vacua are new. We will label these different vacua by an integer related to the centre of the appropriate $SU(N)$ embeddings required for twisting. Hence the integer is taken to be $N$ for the $SU(N)$ based construction, and 6 for the $SU(2) \times SU(3)$ construction. Obviously we reserve the label 1 for the trivial component.

The essence of the $SU(2)$ based construction is the existence of a suitable subgroup $PG_N'$ within a $G_2$ subgroup of $G$. The non-trivial flat connections arise through the decomposition of $G$ into $G_2 \times H$, with $H$ the maximal subgroup commuting with $G_2$. It is the CSA of $H$ that determines the deformations (see the discussion below eq. (\ref{pert})) of the non-trivial $G_2$ connection as embedded in $G$, fixing the dimension of this connected vacuum component (rank($H$)). Since $H$ commutes with the $G_2$ subgroup containing the holonomies, it also plays the role of the maximal unbroken subgroup, apart from some global discrete symmetries. Compare the situation to Witten's D-brane construction \cite{Witnew}.

Our construction presented here goes via the chains
\be
\begin{array}{rrrrrrrrrrr}
E_8 & \rightarrow & E_7 & \rightarrow & E_6 &&&&&&\\
&&&&& \searrow &&&&& \\
&& \multicolumn{3}{r}{SO(2N)} & \rightarrow & SO(8) &&&& \\
&&&&&&& \searrow &&& \\
&&&&&& F_4 & \rightarrow & SO(7) & \rightarrow & G_2 \\
&&&&&&& \nearrow &&& \\
&&&& \multicolumn{3}{r}{SO(2N+1)}&&&&
\end{array}
\ee   
It is this general feature that repeats itself for the $SU(N > 2)$ based constructions. For $SU(3)$ the role of $G_2$ is played by $F_4$, with the embedding chain $E_8 \rightarrow E_7 \rightarrow E_6 \rightarrow F_4$. For $SU(4)$ the role of $G_2$ is played by $E_7$, with chain $E_8 \rightarrow E_7$, and finally for $SU(5)$ and $SU(2) \times SU(3)$, $E_8$ stands alone. Our results are summarised in table (\ref{tab1}), which presents the connected component of the maximal unbroken subgroup for each vacuum-type.

So far we have concentrated on the classical gauge fields. For a calculation of the Witten index $\textrm{Tr}(-)^F$, these should be quantised, and the fermions should be included. The computation is essentially the same as in \cite{IWit, Witnew}: Each vacuum component implies a unique bosonic vacuum, and fermions can be added in $r'+1$ ways, where $r'$ is the dimension of the vacuum component, which is equal to the rank of the unbroken subgroup. Thus each vacuum component contributes $r'+1$ to $\textrm{Tr}(-)^F$. In table \ref{tab2}, we list $r'+1$, where $r'$ is the rank of the group listed in table \ref{tab1}. The columns labeled by $3,4,5,6$ contain more entry's to indicate that there is more than one vacuum component for each type. Finally, table \ref{tab2} also lists the dual Coxeter number for each group. It is easy to verify that, rather miraculously, (\ref{Cox}) is satisfied. 

\section{Conclusions}

We have rederived the results of \cite{Witnew, Keur1}, and embedded these results into the exceptional groups $F_4$ and $E_{6,7,8}$. The basic structure is contained in the exceptional group $G_2$, and the geometry (in a convenient gauge) is basically that of so-called multi-twisted boundary conditions. For $G_2$ and $SO(N)$ our methods are in some sense complementary to those of \cite{Witnew, Keur1}, some aspects are more conveniently described in our construction, other aspects are more transparant in the original approach. For the remaining exceptional groups an approach based on M-theory is not available (yet).

The contributions to $\textrm{Tr}(-)^F$ from the embedding of the $G_2$ non-trivial flat connection in the exceptional groups $F_4$ and $E_{6,7,8}$ is still insufficient to satisfy (\ref{Cox}). The generalisation of our method to $SU(N)^M$ subgroups with $N > 2$ will allow us to derive additional non-trivial flat connections for these exceptional groups. In a subsequent publication \cite{Keur3} we will construct these extra vacua, and demonstrate that they can solve the Witten index problem for the exceptional groups.

{\bf Acknowledgements}: We would like to thank Pierre van Baal, Jan de Boer, Robbert Dijkgraaf, Arkady Vainshtein and Andrei Smilga for helpful discussions.

\appendix

\section{Lie algebras: conventions}
Let $\cal L$ be a Lie algebra. The Killing form is given by:
\be
B(a,b) = \textrm{tr}\{\textbf{ad}(a)\textbf{ad}(b) \}
\ee
A Cartan subalgebra $\cal H$ is a maximal abelian subalgebra in $\cal L$. Because all elements in $\cal H$ commute, they can be simultaneously diagonalised. In particular, in the adjoint representation, the eigenvectors of $\cal H$ are elements of the Lie algebra. One can write:
\be
[ h , e_{\alpha} ] = \alpha(h) e_{\alpha} \quad h \in {\cal H} 
\ee
The $\alpha(h)$ are linear functionals on the space $\cal H$, called roots or root vectors. It is possible to associate elements of $\cal H$ to the functionals $\alpha(h)$ by defining
\be
B(h_{\alpha} , h ) = \alpha(h)
\ee
Because of linearity one has 
\be
h_{p \alpha + q \beta} = p h_{\alpha} + q h_{\beta} 
\ee
The space $\cal H^*$ of linear functionals is a vector space. The roots form a (finite) subset of this space. The set of root vectors will be denoted by $\Delta$.
Because the Killing form is symmetric and bilinear one can introduce the notation
\be
B(h_{\alpha}, h_{\beta}) = \alpha(h_{\beta})= \beta(h_{\alpha}) \equiv \langle \alpha, \beta \rangle 
\ee
For compact Lie algebra's, the Killing form defines an inner product on the root space. The normalisation is determined by a self-consistency condition:
\be
\langle \alpha, \alpha \rangle = B(h_{\alpha}, h_{\alpha}) = \sum_{\gamma \in \Delta} \langle \alpha, \gamma \rangle \langle \gamma, \alpha \rangle
\ee
We now have
\be
[ h_{\alpha}, e_{\beta} ] = \langle \alpha, \beta \rangle e_{\beta}
\ee
The $e_{\beta}$ are normalised such that
\be
[ e_{\alpha}, e_{-\alpha}] = 2 h_{\alpha}
\ee
If $\alpha + \beta \neq 0$:
\be
[ e_{\alpha}, e_{\beta} ] = N_{\alpha, \beta} e_{\alpha + \beta}
\ee
(We will never need an explicit form of $N_{\alpha, \beta}$). Hermitean conjugation acts as follows in our conventions
\be
h_{\alpha}^{\dag} = h_{\alpha} \qquad e_{\alpha}^{\dag}= e_{-\alpha}
\ee
One picks a (non-orthogonal) basis of roots $\alpha_i$ such that, if $\alpha_i$, $\alpha_j$ are in this basis, $\alpha_i - \alpha_j$ is not a root. The roots of such a basis are called "simple". Any root is expressible as $\sum_k c_k \alpha_k$, where the $c_k$ are integers. We always denote simple roots by $\alpha_i$, where $i$ is an index or a number. The simple roots of the compact simple Lie algebra's are listed in appendix B.

Because the elements of the CSA $h_{\alpha}$ always commute, they can be simultaneously diagonalised in any matrix representation. In a specific matrix representation $(h_{\alpha})_{ij}$,  weights $\lambda_i$ are defined by $(h_{\alpha})_{ii} = \langle \alpha, \lambda_i \rangle$ for each $\alpha$. Consequently the number of weights of a representation is equal to its dimension. A weight $\lambda$ of a group is always of the form
\be 
\lambda = \sum_i n_i \Lambda_i + m_i \alpha_i \qquad n_i, m_i \ \in \mathbf{Z}
\ee
where $\Lambda_i$ are the fundamental weights, and $\alpha_i$ the simple roots. The fundamental weights are defined from the simple roots by
\be
2 \frac{ \langle \Lambda_i , \alpha_j \rangle}{\langle \alpha_j , \alpha_j \rangle} = \delta_{ij}
\ee
The fundamental weights are always of the form $\sum_k q_k \alpha_k$ where the $q_k$ are rational numbers.

\section{Lie algebras: roots and weights}

For easy refence we give some quantities for the groups used in this article. Our conventions are basically those of \cite{group}, where they can be found in appendix F. For $G_2$, we find it easier to work with abstract root vectors, for the orthogonal and other exceptional groups it is more convenient to work with explicit forms for the root vectors. We omit $Sp(N)$, since we will never need it for explicit calculations. $E_8$, $F_4$ and $G_2$ do not possess non-integer fundamental weights, and hence none are listed. All fundamental weights of $SU(N)$ are non-integer, but we will not need them, and hence they are not listed either.

We use the notation $e_i$ for the unit vector in the $i$-direction. Inproducts $\langle \alpha , \beta \rangle$ that are not listed can either be obtained by  
 $\langle \alpha , \beta \rangle= \langle \beta , \alpha \rangle$, or are $0$.
In the non-simply laced algebra's, the solid dots in the Dynkin diagrams denote the shorter roots. 

\begin{figure}[!ht]
\begin{center}
\includegraphics[width=8cm]{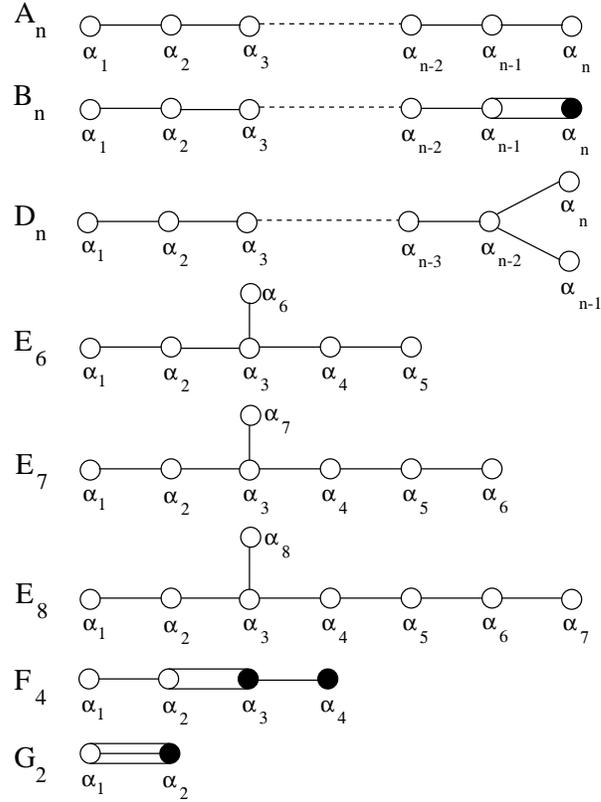}
\end{center}
\caption{Dynkin diagrams}
\end{figure}

\subsection{$SU(n)$ ($A_n$)}
\begin{itemize}
\item Simple roots:
\bd 
\begin{array}{l}
\langle \alpha_i , \alpha_i \rangle = \frac{1}{n+1} \ ( 1 \leq i \leq n) \quad \langle \alpha_i , \alpha_{i + 1} \rangle = - \frac{1}{2(n+1)}  \ ( 1 \leq (i + 1) \leq n)
\end{array}
\ed
\item Positive roots:
\bd
\sum_{p=j}^k \alpha_p \ ( 1 \leq j \leq k \leq n)
\ed
\end{itemize}

\subsection{$SO(2n+1)$ ($B_n$)}
\begin{itemize}
\item Simple roots:
\bd 
\begin{array}{l}
\langle \alpha_i , \alpha_i \rangle = \frac{1}{2n-1} \ (1 \leq i \leq n-1 ) \\ \langle \alpha_n , \alpha_n \rangle = \frac{1}{2(2n-1)} \\
\langle \alpha_i , \alpha_{i + 1} \rangle = - \frac{1}{2(2n-1)}  \ ( 1 \leq (i + 1) \leq n)
\end{array}
\ed
explicit representation:
\bd 
\begin{array}{rcl}
\alpha_i & = & (e_i - e_{i+1}) / \sqrt{2(2n-1)}  \ ( 1 \leq i < n) \\
\alpha_n & = & e_n / \sqrt{2(2n-1)}\\
\end{array}
\ed
\item Positive roots:
\bd
e_k /  \sqrt{2(2n-1)}  \quad (e_k \pm e_l) / \sqrt{2(2n-1)} \ (k < l)
\ed
\item Non-integer fundamental weights:
\bd 
\Lambda_n = \frac{1}{\sqrt{2(2n-1)}} \sum_{i=1}^n e_i/2 
\ed
\end{itemize}

\subsection{$SO(2n)$ ($D_n$)}
\begin{itemize}
\item Simple roots:
\bd 
\begin{array}{l}
\langle \alpha_i , \alpha_i \rangle = \frac{1}{2(n-1)} \ (1 \leq i \leq n) \\ \langle \alpha_i , \alpha_{i + 1} \rangle = - \frac{1}{4(n-1)} \ ( 1 \leq (i + 1) \leq n-3) \\ \langle \alpha_{n-2} , \alpha_{n-1} \rangle = \langle \alpha_{n-2} , \alpha_{n} \rangle = - \frac{1}{4(n-1)} 
\end{array}
\ed
explicit representation:
\bd
\begin{array}{rcl}
\alpha_i & = & (e_i - e_{i+1}) / \sqrt{4(n-1)}  \ ( 1 \leq i < n) \\
\alpha_n & = & (e_{n-1} + e_{n}) / \sqrt{4(n-1)}\\
\end{array}
\ed
\item Positive roots:
\bd
(e_k \pm e_l) / \sqrt{4(n-1)} \ (k < l)
\ed
\item Non-integer fundamental weights:
\bea
\Lambda_{n-1} & = & \frac{1}{\sqrt{2(n-1)}}( \sum_{i=1}^{n-2} e_i/2 - e_{n-1}/2 - e_n /2) \non  \\
\Lambda_{n} & = & \frac{1}{\sqrt{2(n-1)}} ( \sum_{i=1}^{n-2} e_i/2 - e_{n-1}/2 + e_n /2 ) \non 
\eea
\end{itemize}

\subsection{$E_6$}
\begin{itemize}
\item Simple roots:
\bd \begin{array}{l}
\langle \alpha_i , \alpha_i \rangle = \frac{1}{12} \ (1 \leq i \leq 6) \\
\langle \alpha_1 , \alpha_2 \rangle = \langle \alpha_2, \alpha_3 \rangle = \langle \alpha_3, \alpha_4 \rangle = - \frac{1}{24} \\
\langle \alpha_4 , \alpha_5 \rangle  = \langle \alpha_3, \alpha_6 \rangle= - \frac{1}{24}
\end{array}
\ed
explicit representation:
\bd 
\begin{array}{lcr}
\multicolumn{3}{c} {\alpha_1 = (e_1 \sqrt{3} - e_2 - e_3 - e_4 - e_5 - e_6) / 2\sqrt{24} } \\ 
\alpha_2 = (e_5 + e_6) / \sqrt{24} & & \alpha_5 = (e_2 - e_3) / \sqrt{24} \\
\alpha_3 = (e_4 - e_5) / \sqrt{24} & & \alpha_6 = (e_5 - e_6) / \sqrt{24} \\
\alpha_4 = (e_3 -e_4) / \sqrt{24}  & &  
\end{array}
\ed
\item Positive roots:
\bd
(e_k \pm e_l)/ \sqrt{24} \ (1 < k < l) \quad (e_1 \sqrt{3} \pm e_2 \pm e_3 \pm e_4 \pm e_5 \pm e_6) / 2 \sqrt{24}
\ed
In the last expression the number of minus-signs should be odd.
\item Non-integer fundamental weights:
\bd
\begin{array}{ll}
\Lambda_1 = & (\frac{2}{3} \sqrt{3} e_1) / \sqrt{24} \\
\Lambda_2 = & (\frac{5}{6} \sqrt{3}e_1 + \frac{1}{2}(e_2+e_3+e_4+e_5+e_6)) / \sqrt{24} \\
\Lambda_4 = & (\frac{2}{3} \sqrt{3}e_1 + e_2 + e_3) / \sqrt{24}\\
\Lambda_5 = & (\frac{1}{3} \sqrt{3}e_1 + e_2) / \sqrt{24} \\
\end{array}
\ed
\end{itemize}

\subsection{$E_7$}
\begin{itemize}
\item Simple roots:
\bd \begin{array}{l}
\langle \alpha_i , \alpha_i \rangle = \frac{1}{18} \ (1 \leq i \leq 7) \\
\langle \alpha_1 , \alpha_2 \rangle = \langle \alpha_2, \alpha_3 \rangle = \langle \alpha_3, \alpha_4 \rangle = - \frac{1}{36} \\
\langle \alpha_4, \alpha_5 \rangle= \langle \alpha_5 , \alpha_6 \rangle  = \langle \alpha_3, \alpha_7 \rangle= - \frac{1}{36}
\end{array}
\ed
explicit representation:
\bd 
\begin{array}{lcr}
\multicolumn{3}{c}{\alpha_1 = (e_1 \sqrt{2} - e_2 - e_3 - e_4 - e_5 - e_6 - e_7) / 12} \\
\alpha_2 = (e_6 + e_7) / 6 & & \alpha_5 = (e_3 - e_4) / 6 \\
\alpha_3 = (e_5 - e_6) / 6 & & \alpha_6 = (e_2 - e_3) / 6 \\
\alpha_4 = (e_4 - e_5) / 6 & & \alpha_7 = (e_6 - e_7) / 6
\end{array}
\ed
\item Positive roots:
\bd
e_1 /\sqrt{18} \quad (e_k \pm e_l)/ 6 \ (1 < k < l) \quad (e_1 \sqrt{2} \pm e_2 \pm e_3 \pm e_4 \pm e_5 \pm e_6 \pm e_7) / 12
\ed
In the last expression the number of minus-signs should be even.
\item Non-integer fundamental weights:
\bd
\begin{array}{ll}
\Lambda_4 & = (\frac{3}{2}\sqrt{2} e_1 + e_2 + e_3 + e_4)/ 6 \\
\Lambda_6 & = (\frac{1}{2}\sqrt{2} e_1 + e_2)/ 6 \\
\Lambda_7 & = (\sqrt{2}e_1 + \frac{1}{2}(e_2 + e_3 + e_4 + e_5 + e_6 - e_7))/ 6 
\end{array}
\ed
\end{itemize}

\subsection{$E_8$}
\begin{itemize}
\item Simple roots:
\bd \begin{array}{l}
\langle \alpha_i , \alpha_i \rangle = \frac{1}{30} \ ( 1 \leq i \leq 8) \\
\langle \alpha_1 , \alpha_2 \rangle = \langle \alpha_2, \alpha_3 \rangle = \langle \alpha_3, \alpha_4 \rangle = \langle \alpha_4, \alpha_5 \rangle= - \frac{1}{60} \\
\langle \alpha_5 , \alpha_6 \rangle  = \langle \alpha_6, \alpha_7 \rangle = \langle \alpha_3, \alpha_7 \rangle= - \frac{1}{60}
\end{array}
\ed
explicit representation:
\bd 
\begin{array}{lcr}
\multicolumn{3}{c}{ \alpha_1 = (e_1 - e_2 - e_3 - e_4 - e_5 - e_6 - e_7 -e_8) / 2\sqrt{60}} \\
\alpha_2 = (e_7 + e_8) / \sqrt{60} & & \alpha_6 = (e_3 - e_4) / \sqrt{60} \\
\alpha_3 = (e_6 - e_7) / \sqrt{60} & & \alpha_7 = (e_2 - e_3) / \sqrt{60} \\
\alpha_4 = (e_5 - e_6) / \sqrt{60} & & \alpha_8 = (e_7 - e_8) / \sqrt{60} \\
\alpha_5 = (e_4 - e_5) / \sqrt{60} & & 
\end{array}
\ed
\item Positive roots:
\bd
(e_k \pm e_l)/ \sqrt{60} \ (k < l) \quad (e_1 \pm e_2 \pm e_3 \pm e_4 \pm e_5 \pm e_6 \pm e_7 \pm e_8) / 2 \sqrt{60}
\ed
In the last expression the number of minus-signs should be odd.
\end{itemize}

\subsection{$F_4$}
\begin{itemize}
\item Simple roots:
\bd \begin{array}{l}
\langle \alpha_1 , \alpha_1 \rangle = \langle \alpha_2 , \alpha_2 \rangle = \frac{1}{9} \\
\langle \alpha_3 , \alpha_3 \rangle = \langle \alpha_4 , \alpha_4 \rangle = \frac{1}{18} \\
\langle \alpha_1 , \alpha_2 \rangle = \langle \alpha_2, \alpha_3 \rangle = - \frac{1}{18} \\
\langle \alpha_3 , \alpha_4 \rangle = - \frac{1}{36}
\end{array}
\ed
explicit representation:
\bd 
\begin{array}{rcl}
\alpha_1 & = & (e_2 -e_3) / \sqrt{18} \\
\alpha_2 & = & (e_3 -e_4) / \sqrt{18} \\
\alpha_3 & = & (e_4) / \sqrt{18} \\
\alpha_4 & = & (e_1 - e_2 - e_3 - e_4) / 2\sqrt{18} \end{array}
\ed
\item Positive roots:
\bd
e_k/ \sqrt{18} \quad (e_k \pm e_l)/ \sqrt{18} \ (k < l) \quad (e_1 \pm e_2 \pm e_3 \pm e_4) / 2 \sqrt{18}
\ed
\end{itemize}

\subsection{$G_2$}
\begin{itemize}
\item Simple roots:
\bd 
\begin{array}{l}
\langle \alpha_1 , \alpha_1 \rangle = \frac{1}{4} \quad \langle \alpha_1 , \alpha_2 \rangle = - \frac{1}{8} \quad \langle \alpha_2 , \alpha_2 \rangle = \frac{1}{12} 
\end{array}
\ed
\item Positive roots:
\bd
\alpha_1, \alpha_2, \alpha_1 + \alpha_2, \alpha_1 + 2 \alpha_2, \alpha_1 + 3 \alpha_2, 2 \alpha_1 + 3 \alpha_2
\ed
\end{itemize}

\end{document}